\documentclass[12pt,thmsa]{article}

\usepackage{sw20lart}

\usepackage{graphicx}

\usepackage{cite}

\usepackage{dcolumn}

\begin{document}

\title{New Approach to Bounded Quantum--Mechanical Models}
\author{Francisco M. Fern\'{a}ndez \thanks{e--mail: fernande@quimica.unlp.edu.ar} \\
INIFTA (UNLP, CCT La Plata--CONICET), Blvd. 113 y 64 S/N, \\
Sucursal 4, Casilla de Correo 16, 1900 La Plata, Argentina}
\maketitle

\begin{abstract}
We develop an approach for the treatment of one--dimensional bounded
quantum--mechanical
models by straightforward modification of a successful method for
unbounded ones. We apply the new approach to a simple example and show that
it provides solutions to both the bounded and unbounded type of models
simultaneously.
\end{abstract}

\section{Introduction\label{sec:intro}}

The Riccati--Pad\'{e} method (RPM) yields accurate eigenvalues and eigenfunctions
of separable quantum--mechanical models\cite
{FMT89a,FMT89b,F92,FG93,F95c,F95,F95b,F96,F96b,F97,F08,F08b,F08c}. The
approach is based on a rational approximation to a modified logarithmic
derivative of the eigenfunction and the best fit occurs when the eigenvalue
is a root of a Hankel determinant\cite
{FMT89a,FMT89b,F92,FG93,F95c,F95,F95b,F96,F96b,F97,F08,F08b,F08c}.

The roots of the Hankel determinant approach the energies of bound states
and resonances as the determinant dimension increases. The RPM does not
require explicit specification of the boundary condition, the approach commonly
selects the physical one automatically for each problem. The resulting
eigenvalues always correspond to the correct asymptotic behaviour at infinity%
\cite{FMT89a,FMT89b,F92,FG93,F95c,F95,F95b,F96,F96b,F97,F08}. As far as we know
the RPM is the only approach that applies exactly in the same form to both bound
states and resonances.

If the
potential--energy function has poles at two points, then the roots of the
Hankel determinant approach the eigenvalues of the problem with Dirichlet
boundary conditions at such coordinate values\cite{F96,F08b,F08c}. We call
these boundary conditions ``natural''.

In some cases one is interested in that the eigenvalue equation satisfies
``artificial'' bounday conditions. For that reason, in this paper we propose
a modification of the RPM to treat Dirichlet boundary conditions at
arbitrary coordinate locations. In Sec. \ref{sec:method} we introduce the
RPM for one--dimensional models and suggest how to force the desired
boundary conditions. In particular we concentrate on a linear potential that
has proved useful for the treatment of some physical problems. In Sec. \ref
{sec:results} we show results for the chosen eigenvalue equation, and in
Sec. \ref{sec:discussion} we summarize the main features of the RPM and draw
some conclusions.

\section{The method\label{sec:method}}

We introduce the RPM by means of a simple one--dimensional problem of the
form
\begin{equation}
Y^{\prime \prime }(x)+Q(x)Y(x)=0,\;Y(0)=0  \label{eq:dif_Y(x)}
\end{equation}
that depends on an adjustable parameter that is necessary to satisfy the
other boundary condition which we will specify below. For example, in the
case of a dimensionless Schr\"{o}dinger--like equation $Q(x)=V(x)-E$, where
the energy $E$ is the adjustable parameter. For concreteness we restrict to
this case from now on.

In order to apply the RPM we define the modified logarithmic derivative
\begin{equation}
f(x)=\frac{g^{\prime }(x)}{g(x)}-\frac{Y^{\prime }(x)}{Y(x)}  \label{eq:f(x)}
\end{equation}
where the function $g(x)$ is chosen so that $f(x)$ is analytic at $%
x=0$ and therefore can be expanded in a Taylor series
\begin{equation}
f(x)=\sum_{j=0}^{\infty }f_{j}x^{j}  \label{eq:f_series}
\end{equation}
Notice that the coefficients $f_j$ depend on $E$.
The RPM is based on the transformation of the power series (\ref{eq:f_series}%
) into a rational function or Pad\'{e} approximant that satisfies
\begin{equation}
\lbrack M/N](x)=\frac{\sum_{j=0}^{M}a_{j}x^{j}}{\sum_{j=0}^{N}b_{j}x^{j}}%
=\sum_{j=0}^{M+N+1}f_{j}x^{j}+O(x^{M+N+2})  \label{eq:[M/N]}
\end{equation}
where $M=N+d$, $d=0,1,\ldots $. Notice that the rational ansatz has just $%
M+N+1$ adjustable parameters $a_{j}$ and $b_{j}$ to fit the first $M+N+2$
coefficients of the Taylor series (\ref{eq:f_series}). The additional
requirement determines the value of $E$ as a root of the Hankel determinant:
\begin{equation}
H_{D}^{d}(E)=\left| f_{i+j+d+1}(E)\right| _{i,j=0,1,\ldots N}=0,
\label{eq:Hankel}
\end{equation}
where $D=N+1$ is the dimension of the Hankel matrix. Each Hankel determinant
is a polynomial function of $E$ and we expect that there is a sequence of
roots $E^{[D,d]}$, $D=2,3,\ldots $ that converges towards the value of $E$
consistent with the second boundary condition.

Commonly, the Hankel quantization condition (\ref{eq:Hankel}) provides the
eigenvalues consistent with the bound states ($Y(x\rightarrow \infty )=0$)
or the resonances embeded in the continuum (outgoing or incoming waves). The
RPM automatically selects the eigenvalues that are consistent with such
``natural'' boundary conditions\cite
{FMT89a,FMT89b,F92,FG93,F95c,F95,F95b,F96,F96b,F97,F08}.

If the potential--energy function exhibits poles, then the RPM automatically
selects Dirichlet boundary conditions at the corresponding coordinate
points. For example, when $V(x)=V_{0}\sec (x)^{2}$ the RPM selects the
boundary conditions $Y(\pm \pi /2)=0$\cite{F96}, and $Y(\pm R)=0$ when $%
V(x)=a^{2}x^{2}/(1-x^{2}/R^{2})^{2}$\cite{F08b,F08c}.

In some cases one wants to force boundary conditions that are not related to
singular points in the potential--energy function. Suppose that we are
interested in the differential equation (\ref{eq:dif_Y(x)}) with the
boundary conditions $Y(0)=Y(1)=0$. We can force such ``artificial'' boundary
conditions by means of a properly chosen function $g(x)$ in equation (\ref
{eq:f(x)}). In fact, the function $g(x)=x(1-x)$ introduces poles at $x=0$
and $x=1$ into the differential equation for $f(x)$ that we can rewrite as
\begin{equation}
x(1-x)f^{\prime }(x)+2(1-2x)f(x)-x(1-x)f(x)^{2}-x(1-x)Q(x)+2=0
\label{eq:dif_f}
\end{equation}
In this way we expect to obtain the eigenvalues consistent with those
boundary conditions.

For simplicity we consider
\begin{equation}
Q(x)=\epsilon -\lambda x  \label{eq:Q(x)_lin}
\end{equation}
A motivation for this choice is that the resulting differential equation and
boundary conditions are related to a simple model for the study of electrons
in a crystal under the effect of an electric field\cite{RZ71}.
The Schr\"{o}dinger equation
\begin{equation}
-\frac{d^{2}\Phi (X)}{dX^{2}}+eFX\Phi (X)=E\Phi (X),\;\Phi (0)=\Phi (L)=0
\label{eq:Schro_cris}
\end{equation}
provides the states and energy levels of an electron of mass $m$ and charge $%
e$ in a box of impenetrable walls at $X=0$ and $X=L$ (that mimics the finite
size of the crystal) under de effect of an electric field of strength $F$%
\cite{RZ71}. This extremely simple model has also been useful in the study
of the tail of the density of states of a disordered system in the presence
of an electric field\cite{LRS76}. By means of the change of variables $X=Lx$
and $\Phi (Lx)=Y(x)$ one obtains the differential equation (\ref{eq:dif_Y(x)}%
) with the coefficient (\ref{eq:Q(x)_lin}) where $\lambda =2mL^{3}Fe/\hbar
^{2}$, and $\epsilon =2mL^{2}E/\hbar ^{2}$.

Another reason for the choice of such example is that one can write its
solutions exactly in terms of the Airy functions $Ai(z)$ and $Bi(z)$:
\begin{equation}
Y(x)=N\left[ Bi(-\epsilon )Ai(\lambda x-\epsilon )-Ai(-\epsilon )Bi(\lambda
x-\epsilon )\right]  \label{eq:Y(x)_exact}
\end{equation}
where $N$ is a normalization factor, and the dimensionless eigenvalues $%
\epsilon _{n}$, $n=0,1,\ldots $ are given by the quantization condition
\begin{equation}
Bi(-\epsilon )Ai(\lambda -\epsilon )-Ai(-\epsilon )Bi(\lambda -\epsilon )=0
\label{eq:exact_eigenval}
\end{equation}

\section{Results\label{sec:results}}

The application of the RPM is straightforward: we obtain as many
coefficients $f_{j}(\epsilon )$ as necessary from the differential equation
for $f(x)$, construct the Hankel determinants $H_{D}^{d}(\epsilon )$ $%
D=2,3,\ldots $ and calculate their roots. We expect these roots to converge
towards the eigenvalues of the differential equation with the boundary
conditions mentioned above.

Table \ref{tab:eig_bound} shows sequences of roots of the Hankel
determinants that already converge towards the exact eigenvalues given by
equation (\ref{eq:exact_eigenval}) when $\lambda =1$. As in previous
applications of the RPM we appreciate that the rate of convergence of the
Hankel sequences decreases as the energy increases because the denominator
of the rational approximation (\ref{eq:[M/N]}) requires greater values of $N$
to accomodate the increasing number of zeros of the solution $Y(x)$. We
clearly see that this modification of the RPM enables one to solve
eigenvalue equations with ``artificial'' Dirichlet boundary conditions.

\begin{figure}[]
\begin{center}
\includegraphics[width=9cm]{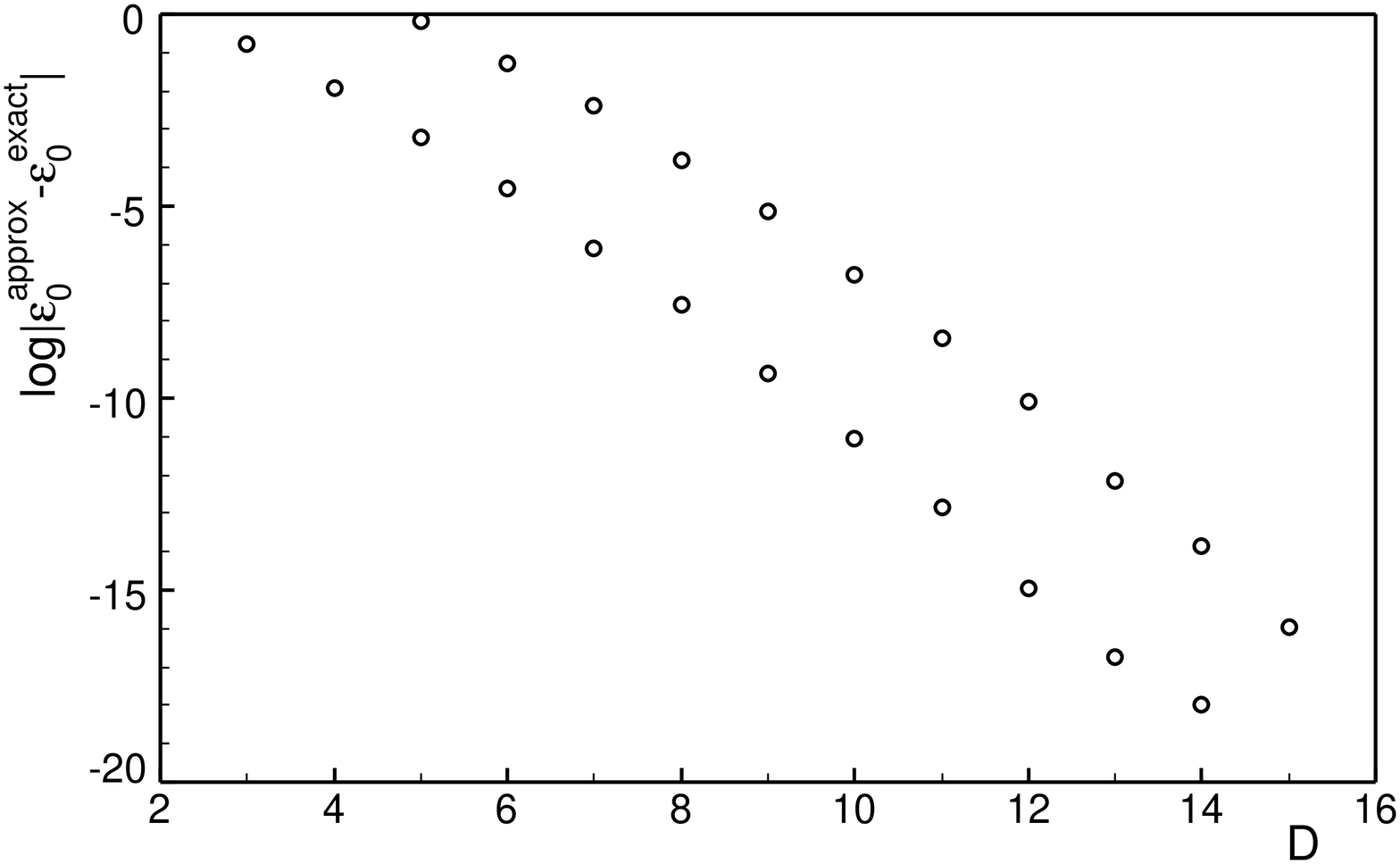}
\end{center}
\caption{Sequences of roots of the Hankel determinants for the lowest eigenvalue
of the bounded model}
\label{Fig:seq_bound}
\end{figure}

The Hankel determinants are polynomial functions of the eigenvalues and
display many more roots than those that we choose to build the sequences
that converge towards the actual eigenvalues of the given problem. One of
the features of the RPM is that an increasing number of roots cluster around
the eigenvalues as $D$ increases. For the simple example chosen here there
are only two roots that approach a given eigenvalue as $D$
increases ( at least for $D\leq 16$). Fig. \ref{Fig:seq_bound} shows $\log
|\epsilon _{0}^{approx}(D)-\epsilon _{0}^{exact}|$ for these two sequences.

\begin{figure}[]
\begin{center}
\includegraphics[width=9cm]{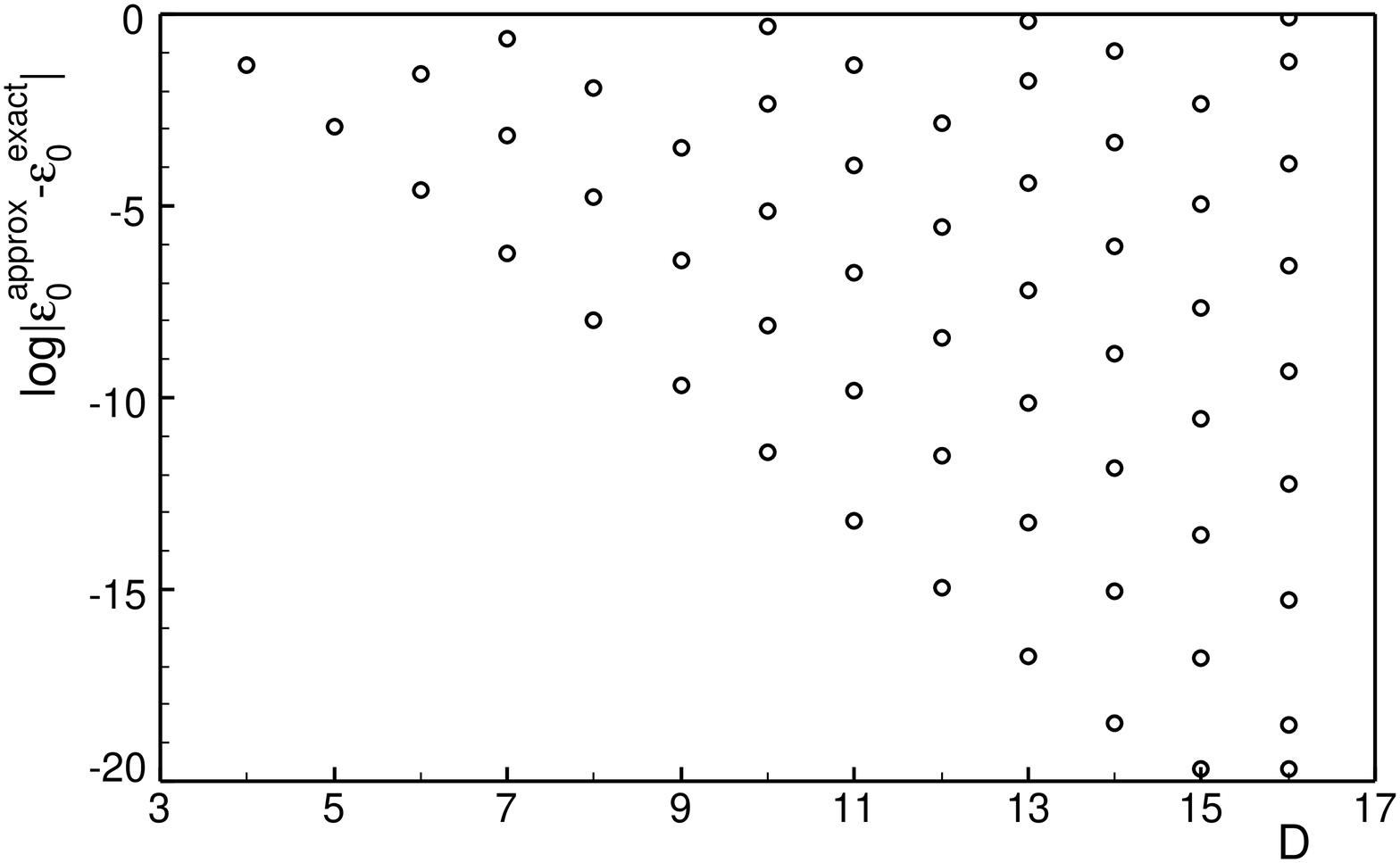}
\end{center}
\caption{Sequences of roots of the Hankel determinants for the lowest eigenvalue
of the unbounded model }
\label{Fig:seq_unbound}
\end{figure}

In the present case the Hankel determinants exhibit other roots than those
mentioned above. They correspond to the ``natural'' boundary condition $%
Y(x\rightarrow \infty )=0$ with eigenvalues given exactly by the quantization
condition $Ai(-\epsilon )=0$. The choice of $g(x)$ suggests that we are
looking for a solution of the form $Y(x)=x(1-x)e^{-\int f(x)\,dx}$ but
the RPM also selects a solution of the form $Y(x)=xe^{-\int
\tilde{f}(x)\,dx}$ with the ``natural'' boundary condition at infinity. The
rational approximation to $\tilde{f}(x)=f(x)+1/(1-x)$ absorbs and removes
the pole at $x=1$ and produces sequences of roots that converge towards the
solutions of the unbounded problem ($0\leq x<\infty $). Table \ref
{tab:eig_unbound} shows some of these eigenvalues for $\lambda =1$.
Curiously, more roots cluster around a given eigenvalue of the unbounded
model than of the bounded one. Fig. \ref{Fig:seq_unbound} shows $\log
|\epsilon _{0}^{approx}(D)-\epsilon _{0}^{exact}|$ for all the sequences
that appear when $D\leq 16$.

\begin{figure}[]
\begin{center}
\includegraphics[width=9cm]{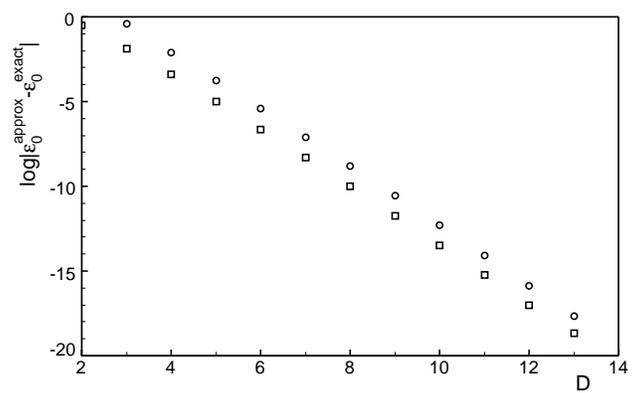}
\end{center}
\caption{Sequences for the lowest eigenvalue of the unbounded model for $%
g(x)=x(1-x)$ (circles) and $g(x)=x$ (squares) }
\label{Fig:seq_unbound2}
\end{figure}

The function $g(x)=x$ is more convenient for the ``natural'' boundary
conditions and, consequently, the sequences of roots of the Hankel
determinants exhibit greater convergence rate. Fig. \ref{Fig:seq_unbound2}
shows $\log |\epsilon _{0}^{approx}(D)-\epsilon _{0}^{exact}|$ for the
optimal sequences for both choices of $g(x)$. When $g(x)=x(1-x)$ the
rational approximation to $\tilde{f}(x)$ has to remove the wrong zero at $x=1
$ and, for this reason, the rate of convergence of the RPM is slightly smaller.
Notice that
the solution $Y(x)$ that satisfies $Y(x)=Y(x\rightarrow \infty )=0$ does not
have a cero at $x=1$.

\section{Discussion\label{sec:discussion}}

Simple models of bounded quantum--mechanical systems have proved useful for
the study of several physical phenomena\cite{F00} (and references therein).
The modification to the RPM proposed here is suitable for bounding a system
between impenetrable walls that force Dirichlet boundary conditions at their
locations. The numerical results of the preceding section show that the
convergence rate of the modified RPM is as remarkable as in the case of the
unbounded and naturally bounded systems\cite
{FMT89a,FMT89b,F92,FG93,F95c,F95,F95b,F96,F96b,F97,F08,F08b,F08c}. A curious
feature of present application of the RPM to a bounded model is that the
approach also provides the eigenvalues of the unbounded one. This outcome is
a consequence of the fact that the RPM automatically selects the correct
asymptotic behaviour at infinity of the solution to the differential
equation. In all the cases studied that asymptotic behaviour coincided with
the one required by physical reasons (vanishing at infinity, incoming or
outcoming waves, etc)\cite
{FMT89a,FMT89b,F92,FG93,F95c,F95,F95b,F96,F96b,F97,F08}.

The transformation of the Schr\"{o}dinger equation into a Riccati one has
proved suitable for the application of the quasilinearization method (QLM)
to quantum mechanics\cite{M99,KM01,LMT06,M06,LDM07,LKM08}. Regarding the
calculation of the resonances of a quartic anharmonic oscillator the RPM\cite{F95}
proves to be more accurate than the QLM\cite{LKM08}.

The main ideas behind the RPM have recently proved useful for the treatment
of two--point nonlinear equations\cite{AF07} of interest in some fields of
physics\cite{BFG07,BBG08,BBG08b}. The resulting approach called Hankel--Pad%
\'{e} method (HPM) appears to be an alternative accurate tool for the
determination of unknown parameters of the theory that are consitent with
the desired asymptotic behaviour of the solution of the nonlinear differential
equation\cite{AF07}.

\begin{table}[tbp]
\caption{First four eigenvalues of the bounded model}
\label{tab:eig_bound}
\begin{center}
\begin{tabular}{D{.}{.}{2}D{.}{.}{20}D{.}{.}{20}}
\hline
\multicolumn{1}{c}{$D$}& \multicolumn{1}{c}{$\epsilon_0$}&
\multicolumn{1}{c}{$\epsilon_1$} \\
\hline

 2  &   9                     &                          \\
 3  &  10.2                   &                          \\
 4  &  10.36                  &                          \\
 5  &  10.3679                &  35                      \\
 6  &  10.36848               &  39.3                    \\
 7  &  10.368506              &  39.89                   \\
 8  &  10.36850713            &  39.97                   \\
 9  &  10.368507161           &  39.978                  \\
10  &  10.368507161827        &  39.9787                 \\
11  &  10.3685071618362       &  39.97874                \\
12  &  10.368507161836336     &  39.9787445              \\
13  &  10.3685071618363371    &  39.97874477             \\
14  &  10.368507161836337126  &  39.9787447892           \\
15  &  10.368507161836337127  &  39.97874478986          \\
16  &  10.368507161836337127  &  39.978744789882         \\
\hline\hline
Exact & 10.368507161836337127  & 39.978744789883354325   \\
\hline\hline

\multicolumn{1}{c}{$D$}&\multicolumn{1}{c}{$\epsilon_2$} &
\multicolumn{1}{c}{$\epsilon_3$} \\
\hline
 8  &  81                    &                           \\
 9  &  88                    &                           \\
10  &  89.1                  &                           \\
11  &  89.3                  &   144                     \\
12  &  89.321                &   156                     \\
13  &  89.3259               &   157.9                   \\
14  &  89.3266               &   158.31                  \\
15  &  89.326628             &   158.39                  \\
16  &  89.3266340            &   158.411                 \\
\hline\hline
Exact& 89.326634542478746080 &   158.41378981431004871  \\
\hline

\end{tabular}
\par
\end{center}
\end{table}

\begin{table}[tbp]
\caption{First four eigenvalues of the unbounded model}
\label{tab:eig_unbound}
\begin{center}
\begin{tabular}{D{.}{.}{2}D{.}{.}{20}D{.}{.}{20}D{.}{.}{20}D{.}{.}{20}}
\hline
\multicolumn{1}{c}{$D$}& \multicolumn{1}{c}{$\epsilon_0$}&
\multicolumn{1}{c}{$\epsilon_1$} \\

\hline

 4  &2.29                   &                        \\
 5  &2.337                  &                        \\
 6  &2.33808                & 4.0                    \\
 7  &2.3381070              & 4.083                  \\
 8  &2.33810740             & 4.0878                 \\
 9  &2.3381074103           & 4.087945               \\
10  &2.338107410456         & 4.0879493              \\
11  &2.33810741045970       & 4.087949441            \\
12  &2.338107410459766      & 4.0879494440           \\
13  &2.33810741045976702    & 4.087949444129         \\
14  &2.3381074104597670382  & 4.08794944413093       \\
15  &2.3381074104597670385  & 4.087949444130970      \\
16  &2.3381074104597670385  & 4.08794944413097060    \\
\hline\hline
Exact &2.3381074104597670385& 4.0879494441309706166  \\
\hline\hline
\multicolumn{1}{c}{$D$}&\multicolumn{1}{c}{$\epsilon_2$} &
\multicolumn{1}{c}{$\epsilon_3$} \\
\hline

 8  &   5.1                    &                         \\
 9  &   5.50                   &                         \\
10  &   5.520                  &                         \\
11  &   5.52054                &  6.74                    \\
12  &   5.5205591              &  6.785                   \\
13  &   5.52055981             &  6.7866                  \\
14  &   5.520559827            &  6.786705                \\
15  &   5.52055982808          &  6.7867080               \\
16  &   5.5205598280950        &  6.786708086             \\
\hline\hline
Exact& 5.5205598280955510591   &  6.7867080900717589988    \\
\hline

\end{tabular}
\par
\end{center}
\end{table}

\end{document}